\documentclass[12pt]{article}
\usepackage{amsmath}
\usepackage{amssymb}
\begin{document}
\vskip 2cm
\begin{center}
{\bf {\Large (Anti-)chiral Superfield
Approach to  Nilpotent Symmetries: Self-Dual Chiral Bosonic Theory }}\\

\vskip 3cm

{\sf N. Srinivas$^{(a)}$, T. Bhanja$^{(a)}$, R. P. Malik$^{(a,b)}$}\\
$^{(a)}$ {\it Physics Department, Centre of Advanced Studies,}\\
{\it Banaras Hindu University, Varanasi - 221 005, (U.P.), India}\\

\vskip 0.1cm


\vskip 0.1cm

$^{(b)}$ {\it DST Centre for Interdisciplinary Mathematical Sciences,}\\
{\it Faculty of Science, Banaras Hindu University, Varanasi - 221 005, India}\\
{\small {\sf {E-mails: seenunamani@gmail.com; tapobroto.bhanja@gmail.com; 
rpmalik1995@gmail.com}}}

\end{center}

\vskip 2.2cm

\noindent
{\bf Abstract:} We exploit the beauty and strength of the symmetry invariant restrictions on the (anti-)chiral 
superfields to derive the Becchi-Rouet-Stora-Tyutin (BRST), anti-BRST and (anti-)co-BRST symmetry transformations 
in the case of a two (1+1)-dimensional (2$D$) self-dual chiral bosonic field theory within the framework
of augmented (anti-)chiral superfield formalism. Our 2$D$ ordinary theory is generalized onto a (2, 2)-dimensional
supermanifold which is parameterized by the superspace variable $Z^M = (x^\mu, \theta, \bar\theta)$
where $x^\mu$ (with $\mu = 0, 1$)  are the ordinary 2$D$ bosonic coordinates and ($\theta,\, \bar\theta$)
are   a pair of Grassmannian variables with their standard relationships:
$\theta^2 = {\bar\theta}^2 =0, \theta\,\bar\theta + \bar\theta\theta = 0$. We impose the 
(anti-)BRST and (anti-)co-BRST invariant restrictions on the (anti-)chiral superfields
(defined on the (anti-)chiral (2, 1)-dimensional super-submanifolds of the above
 {\it general} (2, 2)-dimensional supermanifold) to derive the
above nilpotent symmetries. We do not exploit the mathematical strength of the (dual-)horizontality 
conditions {\it anywhere} in our present investigation. We also discuss the properties of nilpotency, 
absolute anticommutativity and (anti-)BRST and 
(anti-)co-BRST symmetry invariance of the Lagrangian density within the framework of our augmented
(anti-)chiral superfield formalism. Our observation of the absolute anticommutativity property is a completely
{\it novel} result in view of the fact that we have considered {\it only} the (anti-)chiral superfields
 in our present endeavor.\\

\vskip 1cm



\newpage

\section{Introduction}

The model of the 2$D$ self-dual chiral bosonic field theory has found applications in different 
areas of research in theoretical physics, e.g., models of (super)strings, W-gravities, quantum Hall
effect, 2$D$ statistical systems, etc., (see, e.g. [1-12] for details). This model has been 
shown to provide a physical example of Hodge theory [13]. The purpose of our present investigation 
is to apply the augmented version of {\it (anti-)chiral} superfield  approach to BRST formalism [14-17] 
 to discuss this 2$D$ system in a systematic manner.

The usual superfield approach [18-24] to Becchi-Rouet-Stora-Tyutin (BRST) formalism takes
into account the mathematical strength and beauty of the horizontality condition  (HC) to obtain
the off-shell nilpotent and absolutely anticommuting (anti-)BRST symmetry transformations
for the gauge and associated (anti-)ghost fields of a given $p$-form ($p = 1, 2, 3, ...$) gauge theory.
 This has been systematically generalized so as to derive the proper (anti-)BRST
symmetry transformations ($s_{(a)b}$) for the {\it matter}, gauge and (anti-)ghost fields {\it together}
for a given {\it interacting} $p$-form gauge theory (see, e.g. [25-28]). The latter 
superfield formalism exploits the additional restrictions (e.g. gauge invariant restrictions)
which are found to be consistent with the 
celebrated HC. These theoretical formulations have been christened as the augmented version of 
superfield formalism (see, e.g. [25-28]).

In our present investigation, we apply the augmented version of the {\it (anti-)chiral} superfield
formalism 
to the model of 2$D$ self-dual chiral bosonic field theory where we exploit the theoretical power and potential 
of the symmetry invariant restrictions on the (anti-)chiral superfields to derive the proper (anti-)BRST
and (anti-)co-BRST symmetry transformations. 
We do not exploit the mathematical technique of (dual-)HCs {\it anywhere} in our present endeavor. 
This is a {\it novel} feature in our present theoretical method (which thoroughly 
relies on the beauty and strength of the physically motivated symmetry considerations).
Such methods have {\it also} been applied in the context of $\mathcal N = 2$ SUSY quantum mechanical models
[29-32] where the nilpotent $\mathcal N = 2$ SUSY  symmetry transformations have been derived.
From now onwards, we shall be calling our present {\it (anti)-chiral} superfield approach as the
 ``augmented superfield formalism" for the sake of brevity but we shall {\it always} use the (anti-)chiral 
superfields for our further discussions.

To be specific, we wish to mention that, within the framework of the {\it usual} superfield approach to
BRST formalism, a given $D$-dimensional ordinary $p$-form gauge theory is generalized onto a $(D, 2)$-dimensional
supermanifold which is characterized by the superspace coordinates $Z^M = (x^\mu, \theta, \bar\theta)$  where
$x^\mu$ (with $\mu = 0, 1, 2,...D-1$) correspond to the bosonic coordinates and a pair of Grassmannian 
variables $(\theta, \bar\theta)$ satisfy: $\theta^2 = {\bar\theta}^2 =0, \theta\,\bar\theta + \bar\theta\theta = 0$.
The ordinary fields of the gauge theory (which are function of spacetime coordinates $x^\mu$ and defined
on an ordinary Minkowskian spacetime manifold) are generalized to the superfields that are function of the superspace
coordinates $Z^M$ and defined on the $(D, 2)$-dimensional supermanifold. The HC requires that the {\it super} curvature
$(p + 1)$-form should be equated with the {\it ordinary} curvature $(p + 1)$-form. This process of covariant reduction 
leads to the derivation of the proper (i.e. off-shell nilpotent and absolutely anticommuting) (anti-)BRST
symmetry transformations. It turns out that these symmetry transformations (and their corresponding generators)
are identified with the translational generators $(\partial_\theta, \partial_{\bar\theta})$ along the 
Grassmannian directions of the $(D, 2)$-dimensional supermanifold.

Against the backdrop of the above explanations, it is clear that the nilpotency 
$(\partial^2_\theta = \partial^2_{\bar\theta} = 0)$ and the absolute anticommutativity
($\partial_\theta\,\partial_{\bar\theta} + \partial_{\bar\theta}\,\partial_\theta = 0$)
of the translational generators  $(\partial_\theta, \partial_{\bar\theta})$  entail
 upon the (anti-)BRST symmetry transformations ($s_{(a)b}$) and their corresponding generators
($Q_{(a)b}$) to obey the properties of nilpotency ($s^2_{(a)b} = 0, \, Q^2_{(a)b} = 0$) and the 
absolute anticommutativity ($s_b\, s_{ab} + s_{ab} s_b = 0, Q_b\, Q_{ab} + Q_{ab} Q_b = 0$). 
This has been found to be true in the works by Bonora and Tonin [20,21] where the superfields
have been expanded along {\it all} the Grassmannian directions ($1, \theta, \bar\theta, \theta \bar\theta$)
of the $(D, 2)$-dimensional supermanifold. As a consequence, the properties of the
nilpotency and absolute anticommutativity of the (anti-)BRST symmetry transformations (and their
corresponding generators) follow. In our present endeavor, for the sake of simplicity,
 we do {\it not} take the full expansion of the superfields along {\it all} the Grassmannian directions.
As a consequence, our present (anti-)chiral superfield approach to BRST formalism is {\it simple}.

One of the key features of our present method is the use of (anti-)chiral superfields that are defined on the 
(anti-)chiral super-{\it submanifolds}  of the general (2, 2)-dimensional supermanifold on which our 2$D$ self-dual
chiral bosonic theory is generalized. We capture the nilpotency and absolute anticommutativity properties 
of the (anti-)BRST and (anti-)co-BRST symmetries and their generators in our present formalism. 
One of the {\it novel} results of our present endeavor is the
observation that the absolute anticommutativity property of the (anti-)BRST as well as (anti-)co-BRST charges emerge 
very naturally in our present investigation despite the fact that we have taken {\it only} the 
(anti-)chiral super expansions of the (anti-)chiral superfields that are defined on the 
(anti-)chiral super-submanifolds.

Our present endeavor is essential on the following grounds. First and foremost, the symmetry invariant 
restrictions have played a decisive role  in the derivation of the BRST-type symmetries 
in the context of gauge theories (see, e.g. [33,34]) as well as $\mathcal N = 2$ SUSY symmetries of 
the quantum mechanical models [29-32]. We have recently applied this method of derivation in the context 
of 4$D$ Abelian 2-form gauge theory [33] as well as a 1$D$ toy model of a rigid rotor [34]. Thus, it is 
important for us to see its sanctity in the context of 2$D$ model of the self-dual chiral 
bosonic theory which has different dimensionality than 4$D$ and 1$D$. We have
accomplished this goal in our present endeavor. Second, our method of derivation is simple and
physically intuitive where we have  {\it not} used the (dual-) horizontality conditions that are {\it mathematical} in nature.
Finally, our present endeavor is an attempt to provide theoretical richness in the context
of the application of superfield formalism to derive the BRST-type nilpotent symmetries 
for the $p$-form gauge theories (in any arbitrary dimension of spacetime).

Our present paper is organized as follows. First of all, to set-up the notations
and convention, we recapitulate the bare essentials of the (anti-)BRST and (anti-)co-BRST
symmetries of our theory in Sec. 2 within the framework of Lagrangian formulation. Our Sec. 3 
deals with the derivation of (anti-)BRST transformations by exploiting the
(anti-) BRST invariant restrictions on the (anti-)chiral superfields. The subject 
matter of Sec. 4 is connected with the derivation of 
(anti-)co-BRST symmetries from the appropriate restrictions on the (anti-)chiral superfields. 
Sec. 5 is devoted to the discussion of nilpotent (anti-)BRST and (anti-)co-BRST invariance of the Lagrangian 
density within the framework of  augmented superfield formalism. In Sec. 6, we discuss the
properties of nilpotency and absolute anticommutativity of the  (anti-)BRST and (anti-)co-BRST
transformations and corresponding charges. We finally make some concluding remarks in our Sec. 7.

\section{Preliminaries: Lagrangian Formulation}

We begin with the following (anti-)BRST and (anti-)co-BRST invariant Lagrangian density for
the 2$D$ self-dual chiral bosonic field theory (see, e.g. [35] for details)
\begin{eqnarray}
{\cal L}_b &=& \frac {\dot\phi^2}{2} - \frac{\dot v^2}{2} + \dot v\,(v' -\phi') +
 \lambda \,\bigl[\dot\phi -\dot v + v'- \phi' \bigr] 
- \frac{1}{2}\,(\phi' - v')^2\nonumber\\ &+& B\,(\dot\lambda - v - \phi)
+ \frac{B^2}{2} - i\, \dot{\bar C}\, \dot C + 2\, i \,\bar C\, C,
\end{eqnarray}
where the overdot on fields $\phi$ and $v$ (i.e. $\dot \phi = \partial \phi/\partial t,\, \dot v =
 \partial v/\partial t$) 
denotes the derivative w.r.t. the ``time" evolution parameter $t$ and the prime on the
fields (i.e. $\phi'$ and $v'$) stands for the space derivatives
(i.e.  $\phi' = \partial \phi/\partial x,\, v' = \partial v/\partial x$). This establishes the fact that   
$\phi$ and $v$ are function of the spacetime variables ($x$ and $t$). The dynamics of the 2$D$ self-dual
chiral bosonic field theory is such that the evolution parameter $t$ plays a decisive
as well as very special role (see, e.g. [35] for details). In the above, we have $\lambda$ as the 
``gauge" field and $B$ is the Nakanishi-Lautrup type auxiliary field. The fermionic 
($C^2 = \bar C^2 = 0, C\,\bar C + \bar C\,C = 0$) (anti-)ghost fields $(\bar C) C$ are required for the
sake of validity of unitarity in the theory. The gauge-fixing term in the theory is: 
 $-\, \frac{1}{2}\,(\dot\lambda - v - \phi)^2 \, \equiv\,B\,(\dot\lambda - v - \phi) + \frac{B^2}{2} $.

It is straightforward to check that the following  nilpotent ($s_{(a)b}^2 = 0$) 
(anti-)BRST symmetry transformations $s_{(a)b}$ (see, e.g. [13,35])
\begin{eqnarray}
&& s_{ab}\phi = - \bar C,\quad s_{ab} v = - \bar C, \quad s_{ab}\lambda = \dot{\bar C},\quad s_{ab} \bar C = 0,\quad
s_{ab} C = -\,i\,B,\quad s_{ab} B = 0,\nonumber\\ 
&& s_{b}\phi = - C,\,\,\quad s_{b} v = -  C,\,\, \quad s_{b}\lambda = \dot C,
\,\,\quad s_{b} \bar C = +\,i\,B,\quad s_{b} C = 0,\,\, \quad s_{b} B = 0,
\end{eqnarray}
leave the action integral $S = \int d^2x \,{\cal L}_b \equiv \int dx \int dt \, {\cal L}_b$ invariant
because the Lagrangian density ${\cal L}_b$ transforms to the total ``time" derivatives as:
\begin{eqnarray}
s_{ab}\,{\cal L}_b = \frac{\partial}{\partial t}\,\bigl[B\,\dot {\bar C}\bigr],\qquad\qquad \qquad
s_{b}\,{\cal L}_b = \frac{\partial}{\partial t}\,\bigl[B\,\dot C\bigr]. 
\end{eqnarray}
Exploiting the standard tricks and techniques of the Noether theorem, we note that the
above infinitesimal, continuous and nilpotent  $(s_{(a)b}^2 = 0)$ and absolutely anticommuting 
($ s_b\,s_{ab} + s_{ab}\,s_b = 0$) (anti-)BRST symmetry transformations $(s_{(a)b})$ 
lead to the derivation of nilpotent ($Q_{(a)b}^2 = 0$) and absolutely anticommuting 
($ Q_b\,Q_{ab} + Q_{ab}\,Q_b = 0$) conserved charges $Q_{(a)b}$ as follows:
\begin{eqnarray}
Q_{ab} &=&  \int dx\, \bigl[ B\dot {\bar C} - ( \dot\phi - \dot v + v' - \phi' )\, \bar C\bigr]
\,\,\equiv \,\, \int dx \,\bigl[ B\,\dot {\bar C} - \dot B\, \bar C\bigr],\nonumber\\
Q_b &=& \int dx \,\bigl[ B\dot C - ( \dot\phi - \dot v + v' - \phi' )\, C\bigr]
\,\,\equiv\,\,  \int dx\, \bigl[ B\,\dot C - \dot B \,C\bigr].
\end{eqnarray}
The conservation law of the above (anti-)BRST charges ( $Q_{(a)b}$) can be proven by exploiting the following 
Euler-Lagrange (EL) equations of motion (EOM)
\begin{eqnarray}
\dot B &=& \dot\phi - \dot v + v' - {\phi}', \,\,\, B = v + \phi - \dot\lambda,\qquad \ddot C + 2 C = 0,\nonumber\\ 
- B &=& \ddot\phi + \dot\lambda - \dot v' - \lambda' - (\phi'' - v ''),\qquad\qquad \ddot{\bar C} +
 2\,\bar C = 0,\nonumber\\
B &=& \ddot v - 2\,\dot v' + {\dot\phi}' + \dot\lambda - {\lambda}' -({\phi}'' - v''),
\end{eqnarray}
that emerge from the Lagrangian density ${\cal L}_b$  [cf. (1)]. We note that 
$(- 2 B) = \ddot\phi - \ddot v + \dot v' - \dot\phi' \equiv  \ddot B$ from (5) which
implies that $\ddot B + 2 B = 0$ is an off-shoot of the above EL-EOMs which plays an important role
in the proof of  conservation law  (i.e. $\dot Q_{(a)b} = 0$).

Besides the above continuous nilpotent symmetry transformations (2), we have another set of
nilpotent symmetry transformations in the theory. These (anti-)dual-BRST [i.e. (anti-)co-BRST]
symmetry transformations ($s_{(a)d}$) are (see, e.g. [13]):
\begin{eqnarray}
&& s_{ad}\, \lambda = C,\qquad \qquad s_{ad}\, \phi = \frac{1}{2}\,\dot{C},\qquad \qquad s_{ad}\,
v = \,\frac{1}{2}\,{\dot C},
\qquad\qquad s_{ad} \,C = 0, \nonumber\\
&&s_{ad}\, \bar C = \, \frac{i}{2}\,(\dot\phi -\dot v + v' - \phi'),\qquad 
\qquad s_{ad}\,(\dot\phi -\dot v + v' - \phi') = 0,
\qquad s_{ad}\, B = 0,\nonumber\\
&& s_d\, \lambda = \bar C,\,\qquad \qquad s_d\, \phi = \frac{1}{2}\,\dot{\bar C},\,\,\,\qquad 
\qquad s_d \,v = \frac{1}{2}\,\dot{\bar C},
\qquad \qquad \quad s_d\, \bar C = 0,\nonumber\\
&& s_d C =   -\,\frac{i}{2}\,(\dot\phi -\dot v + v' - \phi'), \qquad 
\qquad s_d\,(\dot\phi -\dot v + v' - \phi') = 0,
\qquad s_{d}\, B = 0.
\end{eqnarray}
It can be readily checked that the Lagrangian density transforms to the total
``time" derivatives under the above (anti-)co-BRST symmetry transformations, namely;
\begin{eqnarray}
&&s_{ad}\,{\cal L}_b = \frac{\partial}{\partial t} \,\Bigl[\,\frac{\dot C}{2}\,(\dot\phi -\dot v + v' 
- \phi')\Bigr],\quad \quad
s_d\,{\cal L}_b = \frac{\partial}{\partial t}\, \Bigl[\,\frac{\dot{\bar C}}{2}\,
(\dot\phi -\dot v + v' - \phi')\Bigr].
\end{eqnarray}
As a consequence, the action integral  $S =  \int dx \int dt \,{\cal L}_b$ remains invariant under the
(anti-) co-BRST symmetry transformations, too. Exploiting the basic tenets of Noether theorem, we obtain
the following conserved (anti-)co-BRST [i.e. (anti-)dual-BRST]   charges:
\begin{eqnarray}
Q_{ad} &=& \int dx\,\Bigl[ \frac{\dot{C}}{2} (\dot\phi -\dot v + v' - \phi') - C (\dot\lambda - v - \phi)\,\Bigr]
 \equiv \int dx\, \Bigl[\frac{\dot B\,\dot{C}}{2} + B\,C \Bigr]  \nonumber\\
&\equiv& \frac{1}{2} \int dx\, \Bigl[\dot B\,\dot{C} - \ddot B\, C \Bigr],
\,\,\qquad \Bigl(B = -\, (\dot\lambda - v - \phi),\,\,\qquad \ddot B + 2\,B = 0\Bigr), \nonumber\\
Q_d &=& \int dx\,\Bigl[ \frac{\dot{\bar C}}{2} (\dot\phi -\dot v + v' - \phi') - \bar C (\dot\lambda - v - \phi ) \Bigr]
\,\equiv \int dx \,\Bigl[\frac{\dot B\,\dot{\bar C}}{2} + B\,\bar C \Bigr] \nonumber\\ &\equiv& 
\frac{1}{2} \int dx\, \Bigl[\dot B\,\dot{\bar C} - \ddot B\,\bar C \Bigr],
\,\,\qquad\Bigl(B = -\, (\dot\lambda - v - \phi),\,\,\qquad \ddot B + 2\,B = 0\Bigr).
\end{eqnarray}
The conservation law (i.e. $\dot Q_{(a)d} = 0$) of these charges can be proven by exploiting the EL-EOMs in (5)
which are derived from the Lagrangian density (1). We would like to lay emphasis on the fact that, one of the key
signatures of the (anti-)co-BRST symmetry transformations ($s_{(a)d}$) is the observation that
the gauge-fixing term (i.e. $-\, \frac{1}{2}\,(\dot\lambda - v - \phi)^2$) remains invariant
(i.e. $s_{(a)d}\, \bigl[ \dot\lambda - v - \phi\bigr] = 0$) under them.

In the forthcoming sections, we shall capture all the theoretical ingredients 
(i.e. symmetries, conserved charges, (anti-)BRST and (anti-)co-BRST invariances, etc.), 
discussed in this section, within the framework of augmented version of the {\it (anti-)chiral} 
superfield formalism and provide geometrical meaning to them in the language of specific operators
defined on the (anti-)chiral super-submanifolds.

\noindent
\section{(Anti-)BRST Symmetries: Superfield Approach}

To derive the off-shell nilpotent (anti-)BRST symmetry $s_{(a)b}$  within the framework of superfield
formalism, first of all, we generalize
the 2$D$ fields ($\phi(x),\, v(x), \,\lambda(x),\, C(x),\, \bar C(x)$)
onto the (2, 1)-dimensional (anti-)chiral supermanifolds which are characterized by the superspace 
coordinates ($x^\mu,\, \bar\theta$) and ($x^\mu, \theta$), respectively. For instance, for the 
derivation of the BRST symmetry transformations ($s_b$), we generalize the 2$D$ ordinary fields
to their counterpart anti-chiral superfields on a (2, 1)-dimensional anti-chiral super-submanifold as
 \begin{eqnarray}
&& \phi(x)\,\rightarrow \tilde{\Phi}(x, \bar{\theta})\,=\,\phi(x)\,+\,i\,\bar{\theta}\,f_{1}(x),\nonumber \\
&& v(x)\,\rightarrow \tilde{V}(x, \bar{\theta})\,=\,v(x)\,+\,i\,\bar{\theta}\,f_2(x), \nonumber \\
&& \lambda(x)\, \rightarrow \Lambda(x, \bar{\theta})\,=\,\lambda(x)\,+\,\bar{\theta}\,R(x), \nonumber \\
&& C(x)\,\rightarrow F(x, \bar{\theta})\,=\,C(x)\,+\,i\,\bar{\theta}\,b_1(x), \nonumber \\
&& \bar{C}(x)\,\rightarrow \bar{F}(x, \bar{\theta})\,=\,\bar{C}(x)\,+\,i\,\bar{\theta}\,b_2(x),
\end{eqnarray}
where the set of fields ($f_1(x),\, f_2(x),\, R(x),\,b_1(x),\, b_2(x)\,$), on the right hand side,
are the secondary fields which are to be determined in terms of the basic and auxiliary fields
of the Lagrangian density (1) of our theory. We note that the generalization and super expansion
of the Nakanishi-Lautrup type of auxiliary field $B(x)$ has {\it not} been taken into account because
it remains invariant under the (anti-)BRST and (anti-)co-BRST symmetry transformations 
(i.e. $s_{(a)b}\, B =  s_{(a)d}\, B = 0$). In other words, there is {\it no} (anti-)chiral expansions
for this field which implies that the superfield generalization: 
$B(x) \to \tilde B^{({\cal B},\, AB,\, D,\, AD)} (x, \bar\theta) = B(x)$. 
Here we have taken the results of earlier works [20,21,25-28] where it has been established that the
coefficients of $\theta$ and/or $\bar\theta$  correspond to the nilpotent symmetries.

In this context, it is pertinent to point out that
the BRST invariant quantities (when generalized onto a (2, 1)-dimensional anti-chiral super-submanifold) must
 be independent of the ``soul" coordinate $\bar\theta$ because the latter is {\it not} physically realized.
 This is one of the key and basic tenets of the augmented anti-chiral superfield formalism.
 It is gratifying to state that the following BRST invariant quantities
\begin{eqnarray}
&&s_b \, C = 0, \qquad s_b \, (\phi - v) = 0, \,\,\qquad s_b \, (\lambda\,\dot C ) = 0, 
\,\,\qquad s_b \, (\lambda\, + \dot\phi ) = 0, \nonumber\\
&&s_b \bigl[\dot B\, \lambda\, + i\,\dot{\bar C}\,C \bigr] = 0, \qquad s_b \, (\phi\, C ) = 0, 
\qquad s_b \, \bigl[B\, (\phi + v) -2\,i\, \bar C\, C \bigr] = 0,
\end{eqnarray}
turn out to be useful for us in the BRST invariant restrictions on the anti-chiral superfields
of the (2, 1)-dimensional anti-chiral super-submanifold.

Against the backdrop of the above arguments, it is interesting to note that we have the following
restrictions on the anti-chiral (super)fields of our theory, namely;
\begin{eqnarray}
&& \tilde{\Phi}(x, \bar{\theta}) - \tilde{V}(x, \bar{\theta}) = \phi(x) - v(x), \quad
F^{(\cal B)}(x, \bar{\theta}) = C(x),\quad\tilde{\Phi}(x, \bar{\theta})\,F^{(\cal B)}(x, \bar{\theta})
 = \phi(x)\,C(x), \nonumber \\
&&B(x)\bigl[\tilde{\Phi}(x, \bar{\theta})\,+\,\tilde{V}(x, \bar{\theta})\bigr]-
 2\,i\,\bar{F}(x, \bar{\theta})\,F^{(\cal B)}(x, \bar{\theta}) =
B(x)\bigl[\phi(x)\,+\,v(x)\bigr] - 2\,i\,\bar{C}(x)\,C(x),\nonumber\\
&& \dot{B}(x)\,\Lambda(x, \bar{\theta})\,+\,i\,\dot{\bar{F}}(x, \bar{\theta})\,
\dot{F}^{(\cal B)}(x, \bar{\theta})\,=\,\dot{B}(x)\,\lambda(x)\,+\,i\,
\dot{\bar{C}}(x)\,\dot{C}(x), \nonumber \\
&& \Lambda(x, \bar{\theta})\,\dot{F}^{(\cal B)}(x, \bar{\theta})\,=\,\lambda(x)\,\dot{C}(x), 
\quad\qquad \Lambda(x, \bar{\theta})\,+\,\dot{\tilde{\Phi}}(x, \bar{\theta})\,=\,\lambda(x)\,+\,\dot{\phi}(x),
\end{eqnarray}
where $F^{(\cal B)} (x, \bar{\theta}) = C(x)$  because $s_b\, C = 0$. As a consequence
of the BRST invariance $s_b\, C = 0$, there is no expansion of the superfield 
$F(x, \bar{\theta})$ along $\bar\theta$-direction  (i.e. $b_1(x) = 0$). The superscript
 $(\cal B)$ denotes that the superfield $F(x, \bar\theta)$ has been derived after
 the application of the BRST invariant restriction (i.e. $s_b\, C = 0$).  In addition to this
restriction, we have the following relationships that emerge  from the above BRST invariant restrictions, namely;
\begin{eqnarray}
&& f_1(x)\,=\,f_2(x)\,=\,f(x),\qquad  b_1(x)\,=\,0, \qquad R(x)\,\dot{C}(x)\,=\,0,\quad 
R(x)\,+\,i\,\dot{f}(x)\,=\,0, \nonumber\\ 
&&\dot{B}(x)\,R(x)\,-\,\dot{b_2}(x)\,\dot{C}(x)\,=\,0, \quad 
i\,B(x) f(x)\,+\,b_2(x)\,C(x) = 0, \quad  f(x)\,C(x)\,=\,0.
\end{eqnarray}
The last entry shows that $f(x)$ is proportional to $C(x)$. For the sake of  algebraic
convenience, we choose the non-trivial solution as
$f(x) = i\,C(x)$. This choice, immediately,
entails upon the other secondary fields to be: $ R(x)\,=\,\dot{C}(x),\, b_2(x)\,=\,B(x)$. Thus,
we  have the following expressions for the secondary fields in terms of the basic and auxiliary
fields:
\begin{eqnarray}
R(x)\,=\,\dot{C}(x), \qquad b_2(x)\,=\,B(x), \qquad f(x) = i\,C(x), \qquad b_1(x) = 0.
\end{eqnarray}
We lay emphasis on the fact that our 
choice may differ by an overall constant factor. As a consequence, the expressions in (13)
would {\it also} change accordingly in a consistent manner. We note that this kind of freedom is
{\it also} present in our original continuous symmetry transformations (2).
The substitution of these values in the anti-chiral expansions (9) leads to the following
\begin{eqnarray}
&&\tilde{\Phi}^{(\cal B)}(x, \bar{\theta})\,=\,\phi(x)\,+\,\bar{\theta}\,(-\,C(x))\,\equiv\,
\phi(x)\,+\,\bar{\theta}(s_b\,\phi(x)),\nonumber \\
&&  \tilde{V}^{(\cal B)}(x, \bar{\theta})\,=\,v(x)\,+\,\bar{\theta}\,(-\,C(x))\,\equiv \,
v(x)\,+\,\bar{\theta}\,(s_b\,v(x)), \nonumber \\
&&  \Lambda^{(\cal B)}(x, \bar{\theta})\,=\,\lambda(x)\,+\,\bar{\theta}\,(\dot C(x))\,\equiv\,
\lambda(x)\,+\,\bar{\theta}\,(s_b\,\lambda(x)), \nonumber \\
&& F^{(\cal B)}(x, \bar{\theta})\,=\,C(x)\,+\,\bar{\theta}\,(0)\,\equiv\,
C(x)\,+\,\bar{\theta}\,(s_b\,C(x)), \nonumber \\
&&  \bar{F}^{(\cal B)}(x, \bar{\theta})\,=\,\bar{C}(x)\,+\,\bar{\theta}\,(i\,B)\,\equiv\,
\bar{C}(x)\,+\bar{\theta}\,(s_b\,\bar{C}(x)),
\end{eqnarray}
where the superscript ${(\cal B)}$ stands for the expansion of superfields after the application of
BRST invariant restrictions (11). We note that the coefficient of $\bar\theta$ is nothing but the
BRST symmetry transformations [cf. (2)] in the above expansions.

We now concentrate on the derivation of the anti-BRST symmetry transformations from our augmented
superfield formalism. Towards this goal in mind, first of all, we generalize the ordinary 2$D$ fields 
of the theory onto the superfields defined on the chiral super-submanifold that is characterized 
by the superspace coordinates $(x^{\mu}, \, \theta)$. Thus, we have the following generalizations
and expansions of the chiral superfields, namely;
\begin{eqnarray}
&& \phi(x)\rightarrow \tilde{\Phi}(x,\theta)\,=\,\phi(x)\,+\,i\,\theta\,\bar{f}_1(x), \nonumber \\
&& v(x)\rightarrow \tilde{V}(x, \theta)\,=\,v(x)\,+\,i\,\theta\,\bar{f}_2(x), \nonumber \\
&& \lambda(x)\rightarrow \Lambda(x, \theta)\,=\,\lambda(x)\,+\,\theta\,\bar{R}(x), \nonumber \\
&& C(x) \rightarrow\,F(x, \theta)\,=\,C(x)\,+\,i\,\theta \,\bar{b}_1(x), \nonumber \\
&& \bar{C}(x) \rightarrow\,\bar{F}(x, \theta)\,=\,\bar{C}(x)\,+\,i\,\theta \,\bar{b}_2(x),
\end{eqnarray}
where on the r.h.s., we have the secondary fields 
($\bar{f}_1(x),\, \bar{f}_2(x),\, \bar{R}(x),\, \bar{b}_1(x),\,\bar{b}_2(x)$) which are to be 
determined by exploiting the anti-BRST restrictions. In this context, we note that the following set
of anti-BRST invariant quantities:
\begin{eqnarray}
&&s_{ab} \, (\bar C) = 0, \qquad s_{ab}\,(\phi - v) = 0, \,\,\qquad s_{ab}  \, (\lambda\,\dot{\bar C} ) = 0, 
\,\,\qquad s_{ab}  \, (\lambda\, + \dot\phi ) = 0, \nonumber\\
&&s_{ab}  \bigl[\dot B\, \lambda\, + i\,\dot{\bar C}\,\dot C \bigr] = 0, \qquad s_{ab}  \, (\phi\, C ) = 0, 
\qquad s_{ab} \, \bigl[B\, (\phi + v) - 2\,i\, \bar C\, C \bigr] = 0,
\end{eqnarray}
provides us useful relationships that can be generalized onto the (2, 1)-dimensional chiral
super-submanifold (of the general (2, 2)-dimensional supermanifold).

Against the backdrop of the above logic, we have the following restrictions on the chiral
(super)fields defined on the (2, 1)-dimensional chiral super-submanifold 
\begin{eqnarray}
&&\tilde{\Phi}(x, \theta)\,-\,\tilde{V}(x, \theta)\,=\,\phi(x)\,-\,v(x),\,\,\,\,
 \bar{F}^{(AB)}(x, \theta)\,=\,\bar{C}(x),
\,\,\,\,\Lambda(x, \theta)\,\dot{\bar{F}}(x, \theta)\,=\,\lambda(x)\, \dot{\bar{C}}(x),\nonumber\\
&&\Lambda(x, \theta)\,+\,\dot{\tilde{\Phi}}(x, \theta)\,=\,\lambda(x)\,+\,\dot{\phi}(x), 
\qquad\qquad \tilde{\Phi}(x, \theta)\,\bar{F}^{(AB)}(x, \theta)\,=\,\phi(x)\,\bar{C}(x),\nonumber\\
&& B(x)(\tilde{\Phi}(x, \theta) + \tilde{V}(x, \theta)) - 2\,i\,\bar F^{(AB)}(x, \theta)\,{F}(x, \theta) =
\,B(x)(\phi(x)\,+\,v(x)) - 2\,i\,\bar C(x)\,{C}(x)\,,\nonumber\\
&& \dot{B}(x)\,\Lambda(x, \theta)\,+\,i\,\dot{\bar{F}}^{(AB)}(x, \theta)\,\dot{F}(x, \theta) =\,\dot{B}(x)\,\lambda(x)\,
+\,i\,\dot{\bar{C}}(x)\,\dot{C}(x), 
\end{eqnarray}
where we have taken $\bar{F}^{(AB)}(x, \theta) = \bar{C}(x)$ due to the anti-BRST invariance 
($s_{ab} \, \bar C = 0$). The above anti-BRST restrictions lead to the following useful relationships
amongst the secondary fields and the basic and auxiliary fields of our theory (cf. (1)), namely;
\begin{eqnarray}
&&{\bar f_1}(x)\,=\,{\bar f_2}(x)\,=\,{\bar f}(x),\qquad  {\bar b}_2(x) = 0,\qquad \bar{R}(x)\,+\,i\,\dot{\bar{f}}(x) = 0,
\qquad \bar{f}(x)\,\bar{C}(x) = 0,\nonumber\\
&&\bar R(x)\,\dot{\bar{C}}(x) = 0,\qquad  \dot{B}(x)\,\bar R(x)\,+\,\dot{\bar{b}}_1(x)\,\dot{\bar{C}}(x) = 0, 
\qquad i\,\bar{f}\,B(x)\,-\,\bar{b}_1\,\bar{C}(x) = 0.  
\end{eqnarray}
We have the freedom to choose the expression 
for $\bar R(x)$ in terms of $\dot{\bar C}(x)$ which can differ by an overall constant factor. Accordingly,
other expressions for secondary fields would change in (18). However, this freedom is also present in the 
original transformations (2) where we have the freedom to multiply the transformations by a constant
overall factor (without changing the symmetry property  of our theory).
A non-trivial and simple solution for $\bar{R}(x)$ is: $\bar{R}(x) = \dot{\bar C}(x)$. 
This choice, immediately, implies
that the following
\begin{eqnarray}
\bar b_1(x)\,=\,-\,B(x),\quad\qquad  \bar{f}(x)\,=\,i\,\bar{C}(x),\quad\qquad {\bar b}_2(x) = 0.
\end{eqnarray}
Plugging in these values of the secondary fields into (15) yields
\begin{eqnarray}
&&\tilde{\Phi}^{(AB)}(x, {\theta})\,=\,\phi(x)\,+\,{\theta}\,(-\,\bar C(x))\,\equiv\,
\phi(x)\,+\,{\theta}(s_{ab}\,\phi(x)),\nonumber \\
&&  \tilde{V}^{(AB)}(x, {\theta})\,=\,v(x)\,+\,{\theta}\,(-\,{\bar C}(x))\,\equiv \,
v(x)\,+\,{\theta}\,(s_{ab}\,v(x)), \nonumber \\
&&  \Lambda^{(AB)}(x, {\theta})\,=\,\lambda(x)\,+\,{\theta}\,(\dot {\bar C}(x))\,\equiv\,
\lambda(x)\,+\,{\theta}\,(s_{ab}\,\lambda(x)), \nonumber \\
&& F^{(AB)}(x, {\theta})\,=\,C(x)\,+\,{\theta}\,(-\,i\,B(x))\,\equiv\,
C(x)\,+\,{\theta}\,(s_{ab}\,C(x)), \nonumber \\
&&  \bar{F}^{(AB)}(x, {\theta})\,=\,\bar{C}(x)\,+\,{\theta}\,(0)\,\equiv\,
\bar{C}(x)\,+{\theta}\,(s_{ab}\,\bar{C}(x)),
\end{eqnarray}
where the superscript $(AB)$  on the superfields denotes that these have been obtained after the
application of the anti-BRST invariant restrictions (17). Thus, the coefficient 
of $\theta$, in the above expansions, is nothing but the anti-BRST symmetry transformations (2).

A close look at (14) and (20) implies that there is an intimate relationship between the continuous
(anti-)BRST symmetry transformations (as well as their generators) and the translational generators 
($\partial_\theta, \, \partial_{\bar \theta}$) on the (anti-)chiral super-submanifolds of the
general supermanifold. In other words, we have the following relationships
\begin{eqnarray}
&&\frac{\partial}{\partial\theta}\, \Omega^{(AB)} (x, \,\theta) \,= 
\,s_{ab} \, \omega(x) \,\equiv \,\pm \,i\, \bigl[w(x), \,Q_{ab}\bigr]_{\pm},\nonumber\\
&& \frac{\partial}{\partial\bar\theta} \,\Omega^{({\cal B})} (x, \,\bar\theta) \,= 
\,\,s_{b} \, \omega(x)\, \equiv\, \pm \,i\, \bigl[w(x),\, Q_{b}\bigr]_{\pm},
\end{eqnarray}
where the ($\pm$) signs, as the subscripts on the square bracket, denote the (anti)commutator 
for the generic fields $\omega(x)$ being (fermionic)bosonic in nature.
The above equations demonstrate that the operators ($s_b, \, \partial_{\bar\theta},\, Q_b$) and 
($ s_{ab}, \, \partial_\theta, \, Q_{ab} $ ) are inter-related. Their nilpotency property (i.e. 
$s_{(a)b}^2 = 0, \,\partial_\theta^2 = 0, \partial_{\bar\theta}^2 = 0, \, Q_{(a)b}^2 = 0  $) is
intertwined and there is a geometrical interpretation for the (anti-)BRST symmetry transformations. 
For instance, the relationship in (21) states that the BRST transformations on the generic field $\omega(x)$ 
in the 2$D$ ordinary space is equivalent to the translation of the corresponding superfield 
$\Omega^{(\cal B)} (x, \bar\theta)$  (obtained after exploiting the BRST invariant restrictions [cf. (17)])
along the Grassmannian direction $\bar\theta$ of the (2, 1)-dimensional anti-chiral super-submanifold
(of the general (2, 2)-dimensional supermanifold). In exactly similar fashion, the geometrical interpretation
for the nilpotent ($s^2_{ab} = 0$) anti-BRST symmetry transformations ($s_{ab}$) can be provided.

\section{(Anti-)co-BRST Symmetries: Superfield Formalism}

First of all, we concentrate on the derivation of  dual-BRST (i.e. co-BRST) symmetry
transformations of our theory by exploiting the anti-chiral superfields whose expansions,
along the $\bar\theta$-direction of the (2, 1)-dimensional anti-chiral super-submanifold,
are given in (9). Towards this goal in mind, we note that the following useful and 
interesting set of co-BRST (or dual-BRST) invariant quantities, namely;
\begin{eqnarray}
&&s_{d}\,(\phi - v) = 0,  \qquad s_{d} \, (\bar C) = 0,  \qquad s_{d}  \, (\lambda\,\bar C ) = 0,
\qquad s_{d}\,(\dot\lambda\, - 2\,\phi ) = 0, \,\,\quad s_{d}  \, (\phi\, C ) = 0, \nonumber\\
&&s_{d} \, \bigl[\phi\, (\dot\phi - \dot v + v' - \phi') + i\, \dot{\bar C}\, C \bigr] = 0, 
\qquad \,\, s_{d} \, \bigl[\lambda\, (\dot\phi - \dot v + v' - \phi') + 2\,i\, \bar C\, C \bigr] = 0,
\end{eqnarray}
can be generalized onto the (2, 1)-dimensional anti-chiral super-submanifold and the
following co-BRST invariant restrictions can be imposed on the anti-chiral superfields: 
\begin{eqnarray}
&& \tilde{\Phi}(x, \bar{\theta}) - \tilde{V}(x, \bar{\theta}) = \phi(x) - v(x), \quad
\bar{F}^{(D)}(x, \bar{\theta}) = \bar{C}(x), \quad
\Lambda(x, \bar{\theta})\,\bar{F}^{(D)}(x, \bar{\theta}) = \lambda(x)\,\bar{C}(x), \nonumber \\
&&\tilde{\Phi}(x, \bar{\theta})\,\dot{\bar{F}}^{(D)}(x, \bar{\theta})\,=\,\phi(x)\,\dot{\bar{C}}(x), 
\qquad\qquad \dot{\Lambda}(x, \bar{\theta})\,-\,2\,\tilde{\Phi}(x, \bar{\theta})
\,=\,\dot{\lambda}(x)\,-\,2\,\phi(x), \nonumber \\
&&\Lambda(x, \bar{\theta})\,\bigl[\dot{\tilde{\Phi}}(x, \bar{\theta})\,-\,\dot{\tilde{V}}(x, \bar{\theta})\,+\,
\tilde{V}^\prime(x, \bar{\theta}) \,-\,\tilde{\Phi}^\prime(x, \bar{\theta})\bigr]\,
+\,2\,i\,\bar{F}^{(D)}(x, \bar{\theta})
\,F(x, \bar{\theta}) \nonumber \\
&&~~~~~~~~~~~~~~~~~~~~~=\,\lambda(x)\,\bigl[\dot{\phi}(x)\,-\,\dot{v}(x)\,+\,v^{\prime}(x)\,
-\,\phi^{\prime}(x)\bigr]\,+\,
2\,i\,\bar{C}(x)\,C(x),\nonumber\\
&&\tilde\Phi(x, \bar{\theta}) \,\bigl[\dot{\tilde{\Phi}}(x, \bar{\theta})\,-\,\dot{\tilde{V}}(x, \bar{\theta})\,
+\,\tilde{V}^\prime(x, \bar{\theta}) \,-\,\tilde{\Phi}^\prime(x, \bar{\theta})\bigr]\, 
+ \,i\,\dot{\bar F}^{(D)}(x, \bar{\theta})\,F(x, \bar{\theta}) \nonumber \\
&&~~~~~~~~~~~~~~~~~~~~~=\, \phi(x)\,\bigl[\dot{\phi}(x)\,-\,\dot{v}(x)\,+\,v^{\prime}(x)\,-\,\phi^{\prime}(x)\bigr]\,+\,
i\,\dot{\bar C}(x)\,C(x).
\end{eqnarray}
The above co-BRST invariant restrictions imply that the l.h.s. should, primarily, remain independent
of the $\bar\theta$-coordinate. In other words, we demand that the  co-BRST invariant quantities
(which are physical in some sense) should remain independent of the ``soul" coordinates.
The aforementioned are {\it not} physical objects because they are {\it only} mathematical
artifacts which are used in the superspace formalism. It is worthwhile to mention here that, in the
older literature (see, e.g. [36]), the Grassmannian variables $\theta$ and $\bar\theta$ have 
been christened as the ``soul" coordinates and the bosonic variables $x^\mu (\mu = 0, 1, 2, ..., D-1)$
have been called as the ``body" coordinates.  In the above Eqn. (23), we have taken
${\bar F}^{(D)}(x, \bar\theta) = \bar C(x)$ because the anti-ghost field $\bar C(x)$
remains invariant ($s_d\, \bar C(x) = 0$) under $s_d$. Here we have taken the results from the earlier 
works [20,21] where it has been shown that the coefficients of $\theta$ and $\bar\theta$ in this 
expansion of superfields (along  the Grassmannian directions of the appropriately chosen 
supermanifold) leads to the derivation of {\it nilpotent} symmetries.

The co-BRST invariant restrictions (23) lead to the following relationships between 
the secondary fields and basic fields:
\begin{eqnarray}
&&f_1(x)\,=\,f_2(x)\,=\, f(x),  \qquad  R(x)\,\bar{C}(x)\,=\,0,\,\, \qquad f(x)\,\dot{\bar{C}}(x)\,=\,0, \nonumber\\
&& R(x)\,\bigl[\dot{\phi}(x)\,+\,\dot{v}(x)\,+\,v^{\prime}(x)\,-\,\phi^{\prime}(x)\bigr]\,
+\,2\,\bar{C}(x)\,b_1(x)\,=\,0, \quad \dot{R}(x)\,-\,2\,i\,f(x) = 0, \nonumber\\
&& i\,f(x)\,\bigl[\dot{\phi}(x)\,-\,\dot{v}(x)\,+\,v^{\prime}(x)\,-\phi^{\prime}(x)\bigr]\,
+\,\dot{\bar{C}}(x)\,b_1(x)\,=\,0.
\end{eqnarray}
It is clear that the non-trivial solutions of the restriction $R(x)\,\bar{C}(x)\,=\,0 $ and 
$f(x)\,\dot{\bar{C}}(x)\,=\,0$ are $R(x) \propto \bar C(x)$ and $f(x) \propto \dot{\bar C}(x)$.
However, if we make one of the simplest choices, say: $R(x) = \bar C(x)$, the following 
relationships automatically ensue: 
\begin{eqnarray}
f(x)\,=\,-\frac{i}{2}\,\dot{\bar{C}}(x), 
\qquad  b_1(x)\,=\,-\,\frac{1}{2}\,\bigl[\dot{\phi}(x)\,-\,\dot{v}(x)\,+\,v^{\prime}(x)\,-\,\phi^{\prime}(x)\bigr],
\qquad b_2(x)\,=\,0.
\end{eqnarray}
The substitution of these values of the secondary fields into the expansions (9), leads
to  the following expansions (in terms of the co-BRST transformations $s_d$):
\begin{eqnarray}
&&\tilde{\Phi}^{(D)}(x,\bar\theta)\,=\,\phi(x)\,+\,\bar\theta\,\bigl(\frac{\dot{\bar C}(x)}{2}\Bigr)\,\equiv\,
\phi(x)\,+\,\bar\theta\,\bigl(s_{d}\,\phi(x)\Bigr), \nonumber \\
&&\tilde{V}^{(D)}(x, \bar\theta)\,=\,v(x)\,+\,\bar\theta\,\Bigl(\frac{\dot{\bar C}(x)}{2}
\Bigr)\,\equiv\,v(x)\,+\,\bar\theta
\,\Bigl(s_{d}\,v(x)\Bigr), \nonumber \\
&&\Lambda^{(D)}(x, \bar\theta)\,=\,\lambda(x)\,+\,\bar\theta\,\Bigr(\bar C(x)\Bigr)\,
\equiv\, \lambda(x)\,+\,\bar\theta\,\Bigl(s_{d}\,\lambda(x)\Bigr), \nonumber \\
&&F^{(D)}(x, \bar\theta)\,=\,C(x)\,+\,\bar\theta \,\Bigr[-\,\frac{i}{2}\,\{\dot{\phi}(x)\,-\,\dot{v}(x)\,+
\,v^{\prime}(x)\,-\,\phi^{\prime}(x)\}\Bigr]\nonumber\\ 
&&~~~~~~~~~~~~~~\equiv\,C(x)\,+\,\bar\theta\,\Bigl(s_{d}\,C(x)\Bigr), \nonumber \\
&&\bar{F}^{(D)}(x, \bar\theta)\,=\,\bar{C}(x)\,+\,\bar\theta \, \bigl(0\bigr)
\equiv\,\bar{C}(x)\,+\,\bar\theta\,  \,\Bigl(s_{d}\,\bar{C}(x)\Bigr).
\end{eqnarray}
The superscript $(D)$ on the superfields in (23) and (26) denotes that the above expansions
have been obtained after the application of the co-BRST (or dual-BRST) invariant 
restrictions (23). It is clear that the coefficients of $\bar{\theta}$, on the r.h.s.,
 yield the proper (i.e. off-shell nilpotent) dual-BRST symmetry transformations.

To derive the anti-co-BRST symmetry transformation $(s_{ad})$, we tap the chiral expansion (15) and 
impose the anti-co-BRST invariant restrictions on these superfields. In this context, it can be readily
checked that, we have the validity of the following:
\begin{eqnarray}
&& s_{ad}\,(\phi(x)\,-\,v(x))\,=\,0, \qquad\quad
s_{ad}\,(\lambda(x)\,C(x))\,=\,0, \qquad\quad s_{ad}\,(\dot{\lambda}(x)\,-\,2\,\phi(x))\,=\,0, \nonumber \\
&&s_{ad}\,\bigl[\lambda(x)\,\{\dot{\phi}(x)\,-\,\dot{v}(x)\,+\,v'(x)
\,-\,\phi'(x)\}\,+\,2\,i\,\bar{C}(x)\,C(x)\bigr]\,=\,0,\qquad s_{ad}\,C(x)\,=\,0, \nonumber \\
&& s_{ad}\bigl[\phi(x)\,
\{\dot{\phi}(x)\,-\,\dot{v}(x)\,+\,v'(x)\,-\,\phi'(x)\}\,+\,i\,\bar{C}(x)\,
\dot{C}(x)\bigr]\,=\,0.
\end{eqnarray} 
These anti-co-BRST invariant quantities have to be generalized onto the (2, 1)-dimensional chiral
super-submanifold and the following anti-co-BRST invariant restrictions have to be imposed
on the chiral superfields, namely;
\begin{eqnarray}
&& F^{(AD)}(x, \theta)\,=\,C(x), \qquad\qquad\qquad \tilde{\Phi}(x, \theta)\,-\,\tilde{V}(x,
 \theta)\,=\,\phi(x)\,-\,v(x),
\nonumber \\
&&\Lambda(x, \theta)\,F^{(AD)}(x, \theta)\,=\,\lambda(x)\,C(x),\qquad
\tilde{\Phi}(x, \theta)\,\dot{F}^{(AD)}(x, \theta)\,=\,\phi(x)\,\dot{C}(x), \nonumber \\
&&\dot{\Lambda}(x, \theta)\,-\,2\,\tilde{\Phi}(x, \theta)\,=\,\dot{\lambda}(x)\,-\,2\,\phi(x),
\nonumber \\
&&\tilde{\Phi}(x, \theta)\,\bigl[\dot{\tilde{\Phi}}(x, \theta)\,-\,\dot{\tilde{V}}(x, \theta)\,
+\,\tilde{V}^{\prime}(x, \theta)\,-\,\tilde{\Phi}^{\prime}(x, \theta)\bigr]\,
+\,i\,\bar{F}(x, \theta)\,\dot{F}^{(AD)}(x, \theta)\,\nonumber \\
&&~~~~~~~~~~~~~~~~~~~~=\,\phi(x)\,\bigl[\dot{\phi}(x)\,-\,\dot{v}(x)\,+\,v^{\prime}(x)\,-\,\phi^{\prime}(x)\bigr]\,
+\,i\,{\bar{C}(x)}\,\dot{C}(x)\,, \nonumber \\
&&\Lambda(x, \theta)\,\bigl[\dot{\tilde{\Phi}}(x, \theta)\,-\,\dot{\tilde{V}}(x, \theta)\,
+\,\tilde{V}^{\prime}(x, \theta)\,-\,\tilde{\Phi}^{\prime}(x, \theta)\bigr]\,
+\,2\,i\,{\bar F}(x, \theta)\,{F}^{(AD)}(x, \theta)\nonumber \\
&&~~~~~~~~~~~~~~~~~~~~=\,\lambda(x)\,\bigl[\dot{\phi}(x)\,-\,\dot{v}(x)\,+\,v^{\prime}(x) 
\,-\,\phi^{\prime}(x)\bigr]\,+\,2\,i\,{\bar C}(x)\,C(x),
\end{eqnarray}
where ${F}^{(AD)}(x, \theta) = C(x)$ is due to the anti-co-BRST invariance $s_{ad} \,C(x) = 0$
of the ghost field  $C(x)$. The superscript ($AD$) on superfield $F(x, \theta)$ denotes that 
this superfield has been obtained after the  anti-co-BRST invariant restriction.
The above restrictions (28) lead to the following relationships between the secondary fields and the
basic fields of the theory described by the Lagrangian density (1):
\begin{eqnarray}
&&\bar{b}_1(x)\,=\,0,\qquad \bar{f}_1(x)\,=\,\bar{f}_2\,=\,\bar{f}(x), 
\qquad \bar{R}(x)\,C(x)\,=\,0,\qquad \dot{\bar{R}}(x)\,-\,2\,i\,\bar{f}(x)\,=\,0, \nonumber \\
&&i\,\bar{f}(x)\,\dot{C}(x)\,=\,0,  \qquad i\bar{f}(x)\bigl[\dot{\phi}\,-\,\dot{v}\,+\,v'\,-\phi'\bigr]\,
-\,\dot{C}(x)\,\bar{b}_2(x)\,=\,0, \nonumber \\
&&\bar{R}(x)\bigl[\dot{\phi}(x)\,-\,\dot{v}(x)\,+\,v'(x)\,-\,\phi(x)' \bigr]\,-\,2\,C(x)\,\bar{b}_2(x)\,=\,0.
\end{eqnarray}
A simple choice of $\bar{R}(x)\,=\,C(x)$ immediately implies the following expressions for the secondary fields
 in terms of the basic fields of our theory:
\begin{eqnarray}
\bar{f}(x)\,=\,-\,\frac{i}{2}\,\dot{C}(x), \qquad\qquad \bar{b}_2(x)\,=\,\frac{1}{2}\,\bigl[
\dot{\phi}(x)\,-\,\dot{v}(x)\,+\,v'(x)\,+\,\phi'(x)\bigr].
\end{eqnarray}
Thus, it is evident that the anti-co-BRST invariant restrictions (28) lead to the exact determination of 
the secondary fields in terms of the basic fields.

The substitution of the above values of the secondary fields into the chiral expansions (15)
leads to the following uniform expansions:
\begin{eqnarray}
\tilde{\Phi}^{(AD)}(x,\theta)\,&=&\,\phi(x)\,+\,\theta\,\Bigl(\frac{\dot{C}(x)}{2}\Bigr)\,\equiv\,
\phi(x)\,+\,\theta\,\Bigl(s_{ad}\,\phi(x)\Bigr), \nonumber \\
\tilde{V}^{(AD)}(x, \theta)\,&=&\,v(x)\,+\,\theta\,\Bigl(\frac{\dot{C}(x)}{2}\Bigr)\,\equiv\,v(x)\,+\,\theta
\,\Bigl(s_{ad}\,v(x)\Bigr), \nonumber \\
\Lambda^{(AD)}(x, \theta)\,&=&\,\lambda(x)\,+\,\theta\,\Bigl(C(x)\Bigr)\,
\equiv\, \lambda(x)\,+\,\theta\,\Bigl(s_{ad}\,\lambda(x)\Bigr), \nonumber \\
F^{(AD)}(x, \theta)\,&=&\,C(x)\,+\,\theta \,\bigl(0\bigr)\,\equiv\,C(x)\,
+\,\theta\,\Bigl(s_{ad}\,C(x)\Bigr), \nonumber \\
\bar{F}^{(AD)}(x, \theta)\,&=&\,\bar{C}(x)\,+\,\theta \,\Bigr[\frac{i}{2}\,\{\dot{\phi}(x)\,-\,\dot{v}(x)\,
+\,v^{\prime}(x)\,-\,\phi^{\prime}(x)\}\Bigr]\,
\nonumber\\ &\equiv&\,\bar{C}(x)\,+\,\theta\,\Bigl(s_{ad}\,\bar{C}(x)\Bigr).
\end{eqnarray}
The superscript ($AD$) on the superfields denotes that we have obtained the above expansions after
the application of the anti-dual-BRST (anti-co-BRST) invariant restrictions on the   chiral superfields. 
A close look and careful observation of the expansions (26) and (31) reveals a relationship
between the (anti-)co-BRST symmetry transformation $(s_{(a)d})$ and the translational generators
($\partial_{\theta}$ and $\partial_{\bar{\theta}}$) along the Grassmannian directions of the chiral 
and anti-chiral super submanifolds of the general (2, 2)-dimensional supermanifold. The geometrical
interpretations of $s_{(a)d} $ are similar to the (anti-)BRST symmetry transformations $s_{(a)b} $ which
have already been discussed in our previous section [cf. (21) and discussions after it].
The nilpotency  ($s^2_{(a)d} = 0$) of $s_{(a)d} $ is linked with the nilpotency  ($\partial^2_{\theta} = 0$)
of $\partial_{\theta}$.

\section{Invariance of the Lagrangian density: Superfield Approach and Geometrical Meaning}

We provide the geometrical interpretation for the BRST invariance of the Lagrangian density (1)
within the framework of our augmented superfield formalism. Towards this goal in mind, we generalize
the Lagrangian density (1) (defined on the flat 2$D$ ordinary Minkowskian spacetime manifold)
onto the (2, 1)-dimensional  anti-chiral super-submanifold (of the general (2, 2)-dimensional supermanifold) as: 
\begin{eqnarray}
{\cal L}_b \, \longrightarrow \tilde {\cal L}_b^{(\cal B)} & 
=\,&\frac{1}{2} \bigl[\,\dot{\tilde\Phi}^{(\cal B)}(x, \bar{\theta})
\,\dot{\tilde\Phi}^{(\cal B)}(x, \bar{\theta}) 
- \dot{\tilde V}^{(\cal B)}(x, \bar{\theta})\,\dot{\tilde V}^{(\cal B)}(x, \bar{\theta})\bigr]
\nonumber \\ 
&+& \dot{\tilde V}^{(\cal B)}(x, \bar{\theta})\,\bigl[{\tilde V}^{\prime(\cal B)}(x, \bar{\theta}) - 
{\tilde\Phi}^{\prime(\cal B)}(x, \bar{\theta})\bigr]
- \frac{1}{2}\,\bigl[ {\tilde\Phi'^{(\cal B)}}(x, \bar{\theta}) 
- {\tilde V'^{(\cal B)}}(x, \bar{\theta})\bigr]^2 \nonumber \\
&+& \Lambda^{(\cal B)}(x, \bar{\theta}) \bigl[\dot{\tilde\Phi}^{(\cal B)}(x, \bar{\theta}) - 
\dot{\tilde V}^{(\cal B)}(x, \bar{\theta}) 
+ {\tilde V'^{(\cal B)}}(x, \bar{\theta}) - {\tilde\Phi'^{(\cal B)}}(x, \bar{\theta})\bigr]\nonumber\\
&+& B (x) \,\bigl[ \dot{\Lambda}^{(\cal B)}(x, \bar{\theta}) - {\tilde V^{(\cal B)}}(x, \bar{\theta})
 - \tilde\Phi^{(\cal B)}(x, \bar{\theta}) \bigl] \,+ \,\frac{B^2 (x)}{2}\nonumber \\
&-& \,i\,\dot{\bar F}^{(\cal B)}(x, \bar{\theta}) \,\dot{F}^{(\cal B)}(x, \bar{\theta}) + 
2\,i\, {\bar F}^{(\cal B)}(x, \bar{\theta})\, {F}^{(\cal B)}(x, \bar{\theta}),
\end{eqnarray}
where the superfields are defined on the anti-chiral super-submanifold and  it
 would be noted that $F^{(\cal B)}(x, \bar{\theta}) =  C(x)$. In the above, the superscript 
 (${\cal B}$) stands for the superfields that have been obtained after the application of 
BRST invariant restrictions. 
 Furthermore, we have the following equations: 
\begin{eqnarray}
\tilde{V'}^{(\cal B)}(x, \bar{\theta}) - \tilde{\Phi'}^{(\cal B)}(x, \bar{\theta}) &=& v'(x) - \phi'(x),\nonumber\\
\dot{\tilde{\Phi}}^{(\cal B)}(x, \bar{\theta})\,-\,\dot{\tilde{V}}^{(\cal B)}(x, \bar{\theta}) +
\tilde{V'}^{(\cal B)}(x, \bar{\theta}) - \tilde{\Phi'}^{(\cal B)}(x, \bar{\theta}) &=& 
\dot{\phi}(x) - \dot{v}(x) + v^{\prime}(x) - \phi^{\prime}(x).
\end{eqnarray}
In the generalization: ${\cal L}_b\,\to \,\tilde {\cal L}_b^{(\cal B)}$, we have taken
 the expansion from (14). As a consequence of our earlier result
in (21), it can be checked that the following is true, namely;
\begin{eqnarray}
\frac{\partial}{\partial {\bar{\theta}}}\,\tilde{\cal{L}}_{b}^{(\cal{B})}\,
=\,\frac{d}{dt}\,\bigl[B\,\dot{C}\bigr]\,\equiv\,s_{b}\,{\cal{L}}_b, 
\qquad\qquad \Bigl(\partial_{\bar\theta} \leftrightarrow  s_b\Bigr).
\end{eqnarray}
This relationship provides the geometrical interpretation for the BRST invariance. It states that
this super-Lagrangian density $\tilde{\cal{L}}_{b}^{(\cal{B})} $ is the {\it sum} of composite 
(super)fields that have been obtained after the application of the BRST invariant restriction
(11). This {\it sum} is such that its translation  along the $\bar{\theta}$-direction of the anti-chiral
super-submanifold generates a total ``time" derivative in the ordinary space [(cf. (3), (34)].

Against the backdrop of the discussion about the BRST invariance, we can also talk about the anti-BRST invariance
of the Lagrangian density (1). In this connection, we generalize it on the chiral super-submanifold
 [with $\bar F^{(AB)}(x, \theta) = \bar C(x)$] as:
\begin{eqnarray}
{\cal L}_b\,\to \,\tilde {\cal L}_b^{(AB)} & =\,&\frac{1}{2} 
\bigl[\,\dot{\tilde\Phi}^{(AB)}(x, \theta)\,
\dot{\tilde\Phi}^{(AB)}(x, \theta) - \dot{\tilde V}^{(AB)}(x, \theta)\,
\dot{\tilde V}^{(AB)}(x, \theta)\bigr] \nonumber \\
&+& \dot{\tilde V}^{(AB)}(x, \theta)\,\bigl[{\tilde V}^{{\prime}(AB)}(x, \theta)
- {\tilde\Phi}^{{\prime}(AB)}(x, \theta)\bigr] 
- \frac{1}{2}\,\bigl[ {\tilde\Phi'^{(AB)}}(x, \theta) - {\tilde V'^{(AB)}}(x, \theta)\bigr]^2
\nonumber \\
&+&\Lambda^{(AB)}(x, \theta) \bigl[\dot{\tilde\Phi}^{(AB)}(x, \theta) - \dot{\tilde V}^{(AB)}(x, \theta) 
+ {\tilde V'^{(AB)}}(x, \theta) - {\tilde\Phi'^{(AB)}}(x, \theta)\bigr] \nonumber \\
&+& B (x) \,\bigl[ \dot{\Lambda}^{(AB)}(x, \theta) - {\tilde V^{(AB)}}(x, \theta)
 - \tilde\Phi^{(AB)}(x, \theta) \bigl] \,+ \,\frac{B^2 (x)}{2} \nonumber \\
&-& \,i\,\dot{\bar F}^{(AB)}(x, \theta) \,\dot{F}^{(AB)}(x, \theta) +
 2\,i\, {\bar F}^{(AB)}(x, \theta)\, {F}^{(AB)}(x, \theta),
\end{eqnarray}
where the superscript $(AB)$ on the superfields stands for the superfields that have been obtained
after the application of anti-BRST invariant restrictions [cf. (20)]. It is straightforward to
check that the following is true:
\begin{eqnarray}
\frac{\partial}{\partial{\theta}}\,\tilde{\cal{L}}_{b}^{(\cal{B})}\,
=\,\frac{d}{dt}\,(B\,\dot{\bar{C}})\,\equiv\quad s_{ab}\,{\cal{L}}_b\quad 
\Longleftrightarrow \quad s_{ab} \,\, \leftrightarrow \,\,\partial_\theta.
\end{eqnarray}
Thus, we note that the anti-BRST invariance of the Lagrangian density (1) is captured by the l.h.s.
of the above equation where a derivative ($\partial_\theta$), acting on the super Lagrangian density
$\tilde{\cal{L}}_{b}^{(\cal{B})}$ defined on the (2, 1)-dimensional super-submanifold,
produces the total ``time" derivative in the ordinary space (thereby rendering the action integral
$S\,=\,\int\,d^2x\,{\cal{L}}_b\,\equiv\,\int\,dx\,\int\,dt\,{\cal{L}}_b $ invariant).
Geometrically, this statement is equivalent to the translation of the super Lagrangian density
(35) along the $\theta $-direction of the chiral (2, 1) dimensional super-submanifold 
(of the general (2, 2)-dimensional supermanifold) such that
 this process of translation generates an ordinary total ``time" derivative in the ordinary space.

We now focus on the (anti-)co-BRST invariance of the Lagrangian density ${\cal{L}}_{b} $ within
the framework of augmented superfield formalism. We note that the starting Lagrangian density (1)
can be generalized onto the (2, 1)-dimensional (anti-)chiral super-submanifolds
(of the general (2, 2)-dimensional supermanifold) as:
\begin{eqnarray}
{\cal L}_b\,\to\tilde {\cal L}_b^{(D)} & =\,&\frac{1}{2} \bigl[\,\dot{\tilde\Phi}^{(D)}(x, 
\bar{\theta})\,\dot{\tilde\Phi}^{(D)}(x, \bar{\theta}) 
- \dot{\tilde V}^{(D)}(x, \bar{\theta})\,\dot{\tilde V}^{(D)}(x, \bar{\theta})\bigr]\nonumber \\ 
&+& \dot{\tilde V}^{(D)}(x, \bar{\theta})\,\bigl[{\tilde V}^{{\prime}(D)}(x, \bar{\theta}) - 
{\tilde\Phi}^{{\prime}(D)}(x, \bar{\theta})\bigr]
- \frac{1}{2}\,\bigl[ {\tilde\Phi'^{(D)}}(x, \bar{\theta}) - {\tilde V'^{(D)}}(x, \bar{\theta})\bigr]^2\nonumber\\
&+&\Lambda^{(D)}(x, \bar{\theta}) \bigl[\dot{\tilde\Phi}^{(D)}(x, \bar{\theta}) - \dot{\tilde V}^{(D)}(x, \bar{\theta}) 
+ {\tilde V'^{(D)}}(x, \bar{\theta}) - {\tilde\Phi'^{(D)}}(x, \bar{\theta})\bigr] \nonumber \\
&+& B (x) \,\bigl[ \dot{\Lambda}^{(D)}(x, \bar{\theta}) - {\tilde V^{(D)}}(x, \bar{\theta})
 - \tilde\Phi^{(D)}(x, \bar{\theta}) \bigr] 
+ \,\frac{B^2 (x)}{2} \nonumber \\
&-&\,i\,\dot{\bar F}^{(D)}(x, \bar{\theta}) \,\dot{F}^{(D)}(x, \bar{\theta}) 
+ 2\,i\, {\bar F}^{(D)}(x, \bar{\theta})\, {F}^{(D)}(x, \bar{\theta}),
\end{eqnarray}
\begin{eqnarray}
{\cal L}_b\,\to\tilde {\cal L}_b^{(AD)} & =\,&\frac{1}{2} \bigl[\,\dot{\tilde\Phi}^{(AD)}
(x, \theta)\,\dot{\tilde\Phi}^{(AD)}(x, \theta) 
- \dot{\tilde V}^{(AD)}(x, \theta)\,\dot{\tilde V}^{(AD)}(x, \theta)\bigr] \nonumber \\
&+& \dot{\tilde V}^{(AD)}(x, \theta)\,\bigl[{\tilde V}^{{\prime}(AD)}(x, \theta) - 
{\tilde\Phi}^{{\prime}(AD)}(x, \theta)\bigr]
- \frac{1}{2}\,\bigl[ {\tilde\Phi'^{(AD)}}(x, \theta) - {\tilde V'^{(AD)}}(x, \theta)\bigr]^2 
\nonumber \\
&+&\Lambda^{(AD)}(x, \theta) \bigl[\dot{\tilde\Phi}^{(AD)}(x, \theta) - \dot{\tilde V}^{(AD)}(x, \theta) 
+ {\tilde V'^{(AD)}}(x, \theta) - {\tilde\Phi'^{(AD)}}(x, \theta)\bigr]\nonumber\\
&+& B (x) \,\bigl[ \dot{\Lambda}^{(AD)}(x, \theta) - {\tilde V^{(AD)}}(x, \theta) -
 \tilde\Phi^{(AD)}(x, \theta) \bigl] \,
+\,\frac{B^2 (x)}{2} \nonumber \\
&-& \,i\,\dot{\bar F}^{(AD)}(x, \theta) \,\dot{F}^{(AD)}(x, \theta) 
+ 2\,i\, {\bar F}^{(AD)}(x, \theta)\, {F}^{(AD)}(x, \theta),
\end{eqnarray}
where the superscript $(D)$ and $(AD)$ on the superfields denote the superfields that have been
obtained after the application of the co-BRST and anti-co-BRST invariant
 restrictions. It is self-evident that $\bar F^{(D)}(x, \bar\theta) = \bar C(x)$ 
and $ F^{(AD)}(x, \theta) =  C(x)$ because $s_d\, \bar C(x) = 0$ and  $s_{ad}\, C(x) = 0$.
The (anti-)co-BRST invariance [cf. (7)] of the 2$D$ ordinary Lagrangian density (1) can be captured in the
language of the augmented superfield formalism 
(in terms of the operation of the translational generators) as  follows:
\begin{eqnarray}
&&\frac{\partial}{\partial{\theta}}\,\tilde {\cal L}_b^{(AD)}\,=\,
\frac{\partial}{\partial{t}}\,\Bigl[\frac{\dot{C}}{2}\,(\dot{\phi}\,-\,\dot{v}\,+\,v^{\prime}\,
-\,\phi^{\prime})\Bigr]
\equiv\,s_{ad}\,{\cal {L}}_{b}, \nonumber\\
&&\frac{\partial}{\partial{\bar{\theta}}}\,\tilde {\cal L}_b^{(D)}\,=\,
\frac{\partial}{\partial{t}}\,\Bigl[\frac{\dot{\bar{C}}}{2}\,(\dot{\phi}\,-\,\dot{v}\,+\,v^{\prime}\,
-\,\phi^{\prime})\Bigr]
\equiv\,s_d\,{\cal {L}}_{b}.
\end{eqnarray}
Thus, we find that $\partial_{\theta}\leftrightarrow s_{ad} $ and $\partial_{\bar\theta}\leftrightarrow s_{d}$.
In other words, we note that the (anti-)co-BRST invariance of the Lagrangian density
${\cal L}_b $ [cf. (7)] can be captured within the framework of augmented superfield formalism as 
these translations of the super-Lagrangian densities ${\cal L}^{((A)D)}_b $ along the
$\theta$ and $\bar\theta$ directions of the chiral and anti-chiral super-submanifolds are such that
these translations generate the ordinary ``time" derivatives in the 2$D$ ordinary Minkowskian space thereby rendering
the action integral $S = \int d^2 x\, {\cal L}_b $ invariant for the physically well-defined fields of our
theory which vanish off at infinity.

We end this section  re-emphasizing the fact that the (anti-)BRST and (anti-)co-BRST invariance of the
Lagrangian density (1) in the ordinary 2$D$ flat spacetime manifold  is equivalent to the translation of the corresponding
super-Lagrangian densities [cf. (32), (35), (37), (38)] along the  $(\bar\theta)\,\theta$-directions of the
(anti-)chiral super-submanifolds (of the general (2, 2)-dimensional supermanifold) such that
this process of  translation on {\it these} supermanifolds, generates the ordinary ``time" derivatives.
 This is the geometrical interpretation for the (anti-)BRST and (anti-)co-BRST invariance 
that are present in our theory.

\section{Nilpotency and Absolute Anticommutativity Properties: Superfield Formalism}

We capture here the nilpotency and absolute anticommutativity properties of the conserved (anti-)BRST
and (anti-)co-BRST charges which are the generators for {\it such} symmetry transformations (that have been
christened as (anti-)BRST and (anti-)co-BRST symmetry transformations). It can be readily checked that, 
the following are true, namely;
\begin{eqnarray}
&&Q_{b} = \int dx\, s_{b}\,
\Bigl[ -\,i\,(\bar C\, \dot C - \dot{\bar C}\,C)\Bigr]\, \equiv \, \int dx \,
\Bigl[s_{ab} \, (- \,i\,\dot C \, C)\Bigr],\nonumber\\
&&Q_{ab} = \int dx\, s_{ab}\,\Bigl[ i\,(\bar C\, \dot C - \dot{\bar C}\,C)\Bigr] \,
\equiv \, \int dx \,\Bigl[s_{b} \, (\,i\,\dot{\bar C} \, \bar C)\Bigr],
\end{eqnarray}
where the (anti-)BRST transformations  ($s_{(a)b}$)  are given in (2) and the expressions for the 
(anti-)BRST charges are quoted in (4). In view of the mappings: 
$s_b \leftrightarrow \partial_{\bar\theta}, \, s_{ab} \leftrightarrow \partial_{\theta}$, 
we can express (40) in terms of the superfields (with $F^{(\cal B)}(x, \bar\theta) = C(x), 
\,\bar F^{(AB)}(x, \theta) = \bar C(x)$) as follows:
\begin{eqnarray}
Q_{b} &=& -\,i\, \int dx \Bigl[\frac{\partial}{\partial\bar\theta}\,
(\bar F^{({\cal B})}(x, \bar{\theta})\, \dot F^{(\cal B)}(x, \bar{\theta})
- \dot {\bar F}^{(\cal B)}(x, \bar{\theta})\,  F^{({\cal B})}(x, \bar{\theta}))\, \Bigr] \nonumber\\
&\equiv & -\,i\,\int dx \Bigl[\int d\bar\theta \, (\bar F^{({\cal B})}(x, \bar{\theta})\, 
\dot F^{(\cal B)}(x, \bar{\theta})
- \dot {\bar F}^{(\cal B)}(x, \bar{\theta})\,  F^{({\cal B})}(x, \bar{\theta}))\Bigr],\nonumber\\
Q_{b} &=& -\,i\,\int dx \Bigl[\frac{\partial}{\partial\theta}\,({\dot F}^{(AB)}(x, \theta)\,
 F^{(AB)}(x, \theta))\, \Bigr] \nonumber \\
&\equiv &  -\,i\,\int dx \Bigl[\,\int \, d\theta\,( {\dot F}^{(AB)}(x, \theta)\, 
F^{(AB)}(x, \theta))\, \Bigr],\nonumber\\
Q_{ab} &=& \,i\, \int dx \Bigl[\frac{\partial}{\partial\theta}\,( \bar F^{(AB)}(x, \theta)\, \dot F^{(AB)}(x, \theta)
- \dot {\bar F}^{(AB)}(x, \theta) \, F^{(AB)}(x, \theta) ) \Bigr] \nonumber\\
&\equiv & \,i\,\int dx \Bigl[\int d\theta \, (\bar F^{(AB)}(x, \theta)\, \dot F^{(AB)}(x, \theta)
- \dot {\bar F}^{(AB)}(x, \theta) \, F^{(AB)}(x, \theta) )\Bigr],\nonumber\\
Q_{ab} &=& \,i \int dx \Bigl[\frac{\partial}{\partial{\bar\theta}}\,(\dot{\bar F}^{({\cal B})}(x, \bar{\theta})\, 
\bar F^{({\cal B})}(x, \bar{\theta}))\Bigr] 
\equiv  i\int dx \Bigl[\int  d\bar\theta\,(\dot {\bar F}^{({\cal B})}(x, \bar{\theta})\,
 {\bar F}^{({\cal B})}(x, \bar{\theta}))\Bigr].
\end{eqnarray}
The above equations immediately imply that $\partial_{\bar\theta} \, Q_b = 0$ due to the nilpotency
($\partial^2_{\bar\theta} = 0$) of  $\partial_{\bar\theta}$. In exactly similar fashion, we note that 
$\partial_{\theta} \, Q_{ab} = 0$  due to the nilpotency ($\partial^2_{\theta} = 0$) of
the translational generator   $\partial_{\theta}$. However, when we translate the above observations
in the language of ordinary (anti-)co-BRST symmetry transformations and charges, we have:
\begin{eqnarray}
&&\partial_{\bar\theta} \, Q_b = 0 \,\,\quad \Longleftrightarrow\quad\,\, s_b\, Q_b = i\,\{ Q_b, \, Q_{b}\} = 0 \,
\,\,\quad\,\,\,\, \Longrightarrow \quad  Q^2_b = 0, \nonumber\\
&&\partial_{\theta} \, Q_{ab} = 0 \quad \Longleftrightarrow \quad s_{ab}\, Q_{ab} = 
i\,\{Q_{ab}, \, Q_{ab}\} = 0 \quad \Longrightarrow \quad Q^2_{ab} = 0.
\end{eqnarray}
Thus, we observe that the nilpotency ($Q^2_{(a)b} = 0$) of the (anti-)BRST charges $Q_{(a)b}$ is connected with the 
nilpotency ($\partial^2_{\theta} = \partial^2_{\bar\theta} = 0$) of the translational generators
($\partial_{\theta}, \partial_{\bar\theta}$) along the (anti-)chiral
super-submanifolds of the general (2, 2)-dimensional supermanifold.

It is very interesting to observe that $\partial_{\theta} \,Q_b = 0$ and 
$\partial_{\bar\theta} \, Q_{ab} = 0$, too. This can be translated into the ordinary 2$D$ space 
(with  $s_b \leftrightarrow \partial_{\bar\theta}, \, s_{ab} \leftrightarrow \partial_{\theta}$) as follows:
\begin{eqnarray}
&&\partial_{\theta} \, Q_b = 0 \,\,\quad \Longleftrightarrow \quad s_{ab}\, Q_b = i\,\{Q_b, \, Q_{ab}\} = 0 \quad
\Longrightarrow \quad Q_b \, Q_{ab} + Q_{ab}\, Q_b = 0,  \nonumber\\
&&\partial_{\bar\theta} \, Q_{ab} = 0 \quad \Longleftrightarrow \quad s_{b}\, Q_{ab} = i\,\{Q_{ab}, \, Q_b\} = 0 \quad
\Longrightarrow \quad  Q_{ab}\, Q_b + Q_b \, Q_{ab} = 0.
\end{eqnarray}
Thus, the nilpotency and absolute anticommutativity of the (anti-)BRST charges can be 
captured in the language of augmented superfield formalism. In other words, the nilpotency 
($\partial^2_{\theta} = \partial^2_{\bar\theta} = 0$) of the translational 
generators as well as absolute anticommutativity 
($\partial_{\theta} \, \partial_{\bar\theta} + \partial_{\bar\theta}\, \partial_{\theta} = 0$) 
of the translational generators ($\partial_{\theta}, \partial_{\bar\theta}$)
on the (anti-)chiral super-submanifolds are intimately connected with {\it such}
properties associated with the (anti-)BRST charges $Q_{(a)b}$ (and the (anti-)BRST 
symmetry transformations $s_{(a)b}$ they generate). The symmetries  $s_{(a)b}$
 also satisfy: $s^2_{(a)b} = 0$ as well as $s_b \, s_{ab} + s_{ab}\, s_b = 0$ 
which are nothing but the nilpotency and absolute anticommutativity properties 
of the (anti-)BRST symmetry transformations.

We now concentrate on the nilpotency and absolute anticommutativity properties of the 
(anti-)co-BRST charges $Q_{(a)d}$ and the symmetry translations they generate. It can be checked 
that the conserved (anti-)co-BRST charges can be expressed as: 
\begin{eqnarray}
&&Q_d = \,i\, \int dx \,s_d\, \bigl[C\, \dot{\bar C} + \bar C \, \dot C \bigr]
\equiv \,i\, \int dx \,s_{ad} \,\bigl (\dot {\bar C}\, \bar C \bigr),\nonumber\\
&&Q_{ad} = i\, \int dx \,s_{ad} \,\bigl[C\, \dot{\bar C} + \bar C \, \dot C \bigr]
\equiv -\,i\, \int dx\, s_d \, \bigl (\dot C\, C \bigr).
\end{eqnarray}
The above expressions  for the (anti-)co-BRST charges $Q_{(a)d}$ can be also
expressed in terms of the superfields, Grassmannian differentials and derivatives in 
view of the general relationship: 
$s_d \leftrightarrow \partial_{\bar\theta}, \, s_{ad} \leftrightarrow \partial_{\theta}$. 
In other words, we have the following expressions for the (anti-)co-BRST charges  
$Q_{(a)d}$ within the framework of augmented superfield formalism:
\begin{eqnarray}
Q_d &=& \int dx \,\Bigl[\frac{\partial}{\partial{\bar\theta}}\,(i\, \bar F^{(D)}(x, \bar{\theta})\,
\dot F^{(D)}(x, \bar{\theta})
- i\, \dot{\bar F}^{(D)}(x, \bar{\theta})\,F^{(D)}(x, \bar{\theta}))\,\Bigr]\nonumber\\
&\equiv & \int dx \Bigl[\int d\bar\theta\,(i \bar F^{(D)}(x, \bar{\theta})\dot F^{(D)}(x, \bar{\theta}) 
- i\, \dot{\bar F}^{(D)}(x, \bar{\theta})\,F^{(D)}(x, \bar{\theta}))\,\Bigr],\nonumber\\
Q_{d} &=& \int dx \,\Bigl[\frac{\partial}{\partial{\theta}}\,(i\, 
\dot{\bar F}^{(AD)}(x, \theta)\,{\bar F}^{(AD)}(x, \theta))\,\Bigr] \nonumber \\
&\equiv &  \int dx \Bigl[\int d\theta\,(i\, \dot{\bar F}^{(AD)}(x, \theta)\,{\bar F}^{(AD)}
(x, \theta))\,\Bigr],\nonumber\\
Q_{ad} &=& \int dx \Bigl[\frac{\partial}{\partial\theta}\,(i \bar F^{(AD)}(x, \theta)\, 
\dot F^{(AD)}(x, \theta) 
- i\,\dot{\bar F} ^{(AD)}(x, \theta))\, F^{(AD)}(x, \theta) \, \Bigr] \nonumber\\
&\equiv & \int dx \Bigl[\int d\theta \, (i\bar{F}^{(AD)}(x, \theta)\dot F^{(AD)}(x, \theta) 
- i\,\dot{\bar F}^{(AD)}(x, \theta))\, F^{(AD)}(x, \theta)\Bigr],\nonumber\\
Q_{ad} &=& -\,i\,\int dx \Bigl[\frac{\partial}{\partial\bar\theta}\,
(\dot F^{(D)}(x, \bar{\theta})\, F^{(D)}(x, \bar{\theta}))\, \Bigr] \nonumber \\
&\equiv &  -\,i\,\int dx \Bigl[\int\,d\bar \theta\,( \dot F^{(D)}(x, \bar{\theta})\,
F^{(D)}(x, \bar{\theta}))\, \Bigr],
\end{eqnarray}
where, as is evident from our earlier discussions, we have taken ${\bar F}^{(D)}(x, \bar\theta) = \bar C(x),$
${F}^{(AD)}(x, \theta) = C(x)$ due to $s_d\, \bar C(x) = 0$ and $s_{ad}\, C(x) = 0$.
Thus, we note that, as is the case with the (anti-)BRST charges,  there are  {\it two}
types of representations for the conserved (anti-)co-BRST charges
$Q_{(a)d}$ in the language of superspace coordinates, superfields, Grassmannian differentials and 
derivatives, too.

The nilpotency and absolute anticommutativity of the (ant-)co-BRST charges $Q_{(a)d}$ are hidden
in the above expressions for these charges within the framework of superfield formalism. For instance,
it can be seen that, the following is true, namely;
\begin{eqnarray}
&&\partial_{\bar\theta} \, Q_d = 0 \,\,\quad \Longleftrightarrow\quad\,\, s_d\, Q_d = i\,\{Q_d, \, Q_{d}\}\,\,\,\,
\,\,\,\quad \Longrightarrow \quad  Q^2_d = 0, \nonumber\\
&&\partial_{\theta} \, Q_{ad} = 0 \quad \Longleftrightarrow \quad s_{ad}\, Q_{ad} = i\,\{Q_{ad}, \, Q_{ad}\}
\quad \Longrightarrow \quad Q^2_{ad} = 0.
\end{eqnarray}
The above equations demonstrate that the nilpotency ($Q^2_{(a)d} = 0$) of $Q_{(a)d}$ and nilpotency of 
($\partial^2_{\theta} = \partial^2_{\bar\theta} = 0$) of the translational generators
($\partial_{\theta}, \partial_{\bar\theta}$) are inter-connected. Furthermore, we note that the 
alternative expression for $Q_{(a)d}$, in the above equations, are such that:
\begin{eqnarray}
&&\partial_{\theta} \, Q_d = 0\,\,\, \quad \Longleftrightarrow \quad s_{ad}\, Q_d = i\,\{Q_d, \, Q_{ad}\} = 0 \quad
\Longrightarrow \quad Q_d \, Q_{ad} + Q_{ad}\, Q_d = 0,  \nonumber\\
&&\partial_{\bar\theta} \, Q_{ad} = 0 \,\quad \Longleftrightarrow \quad s_{d}\, Q_{ad} = i\,\{Q_{ad}, \, Q_d\} = 0 \quad
\Longrightarrow \quad  Q_{ad}\, Q_d + Q_d \, Q_{ad} = 0,
\end{eqnarray}
which demonstrate that the nilpotency ($\partial^2_{\theta} = \partial^2_{\bar\theta} = 0$) 
of the translational generators 
($\partial_{\theta}$ and  $\partial_{\bar\theta}$) is also connected with the absolute anticommutativity 
($\partial_{\theta} \, \partial_{\bar\theta} + \partial_{\bar\theta}\, \partial_{\theta} = 0$) 
properties. This is due to the fact that the nilpotency property ($\partial^2_{\theta} = \partial^2_{\bar\theta} = 0$)
is the limiting case of the absolute anticommutativity 
($\partial_{\theta} \, \partial_{\bar\theta} + \partial_{\bar\theta}\, \partial_{\theta} = 0$)
when $\partial_{\theta} = \partial_{\bar\theta}$ (or $\partial_{\bar\theta} = \partial_{\theta}$).
This is the reason that we see that, in the ordinary 2$D$ system,
we have the validity of  relations: 
$Q^2_{(a)d} = 0$ and $Q_{ad}\, Q_d + Q_d \, Q_{ad} = 0$. We conclude that the nilpotency and absolute anticommutativity
properties of the (anti-)co-BRST charges (and the continuous symmetries $s_{(a)d}$ they generate)
are intimately connected 
with such properties associated with the translational generators ($\partial_\theta, \partial_{\bar\theta}$) along
the Grassmannian  directions of the (2, 1)-dimensional chiral and anti-chiral super-submanifolds of 
the general (2, 2)-dimensional
supermanifold (on which our present 2$D$ self-dual chiral bosonic field theory is generalized).

\section{Conclusions}

The central theme of our present investigation has been to provide theoretical richness in the realm 
of superfield approach to BRST formalism. We have exploited the potential and power of the symmetry
invariant restrictions on the (super)fields to derive the nilpotent and absolutely anticommuting (anti-)BRST and
(anti-)co-BRST symmetries. We have also provided the geometrical basis for the symmetry transformations,
nilpotency, absolute anticommutativity and symmetry invariance of the Lagrangian density of the 2$D$ self-dual
chiral bosonic field theory (within the framework of augmented superfield formalism).

One of the key results of our present investigation has been the observation that we do {\it not} need
the {\it mathematical} strength of the (dual-)horizontality conditions to derive the {\it proper} 
(anti-)BRST and (anti-)co-BRST symmetry transformations. Further, we have found that the geometrical interpretation
for the symmetry transformations remain intact in our present method of derivation
 {\it vis-{\`a}-vis} the {\it ones} where 
the mathematical strength of the (dual-)horizontality conditions is properly utilized [37]. Thus, both methods of 
derivation are consistent and, in some cases, complementary to each-other (see, e.g. [20,21,33,34]).

The other key result of our present endeavor is the observation that the nilpotency
($Q^2_{(a)b} = 0,\, Q^2_{(a)d} = 0$) and  absolute anticommutativity 
($Q_b \, Q_{ab} + Q_{ab}\, Q_b  = 0, \,Q_d \, Q_{ad} + Q_{ad}\, Q_d  = 0$) of the 
(anti-)BRST  and (anti-)co-BRST charges $Q_{(a)b}$ and $Q_{(a)d}$ are captured within our present 
method of superfield formalism where we have considered {\it only} the (anti-)chiral 
superfields for our whole discussion. This is a {\it novel} result in our previous [33,34] 
and present investigations. These {\it two} sacrosanct properties emerge {\it automatically}
in the superfield approach where the {\it full} expansions of superfields along {\it all}
the Grassmannian directions ($1,\, \theta,\, \bar\theta,\, \theta\bar\theta$) of the
supermanifolds are taken into account (see, e.g. [20,21]).

One of the shortcomings of our present method of superfield formalism is the fact that
it is valid {\it only} for those systems where the Curci-Ferrari (CF) conditions [38] 
turn out to be {\it trivial} so that there exist a {\it single} (anti-)BRST and
(anti-)co-BRST invariant Lagrangian density for the system and nilpotency and absolute 
anticommutativity of the (anti-)BRST and (anti-)co-BRST symmetry transformations are
satisfied in a straightforward manner. The 2$D$ self-dual chiral bosonic field theory 
(which is the topic of discussion for our present endeavor) belongs to {\it this} 
category. Hence, our method is applicable   (as the CF-type condition is {\it trivial} here).
 In the cases where the CF-conditions are {\it non-trivial},
 it is the {\it full} expansions of the superfields that are applicable because the 
 CF-type conditions as well as the derivation of the proper (anti-)BRST and
(anti-)co-BRST symmetries result in very naturally in such kind of superfield formalism.
 We emphasize that, in the latter case, the (dual-)horizontality conditions are necessarily exploited, too.

It would be a nice future endeavor for us to check the sanctity of our theoretical method
in the context of other physical systems of interest. In particular, we are very much
eager to apply our method for the BRST description of the  SUSY gauge theories which are important
because of their relevance (i) in the context of search for the existence of the SUSY particles, and (ii) 
as the field theoretic limiting cases of the (super)string theories. We have successfully
applied our method for the description of $\mathcal{N} = 2 $ SUSY quantum mechanical
models [29-32] but the challenge is to apply our method in the description of 
$\mathcal{N} = 2, 4, 8$ SUSY gauge theories. We are currently busy with these ideas 
and we plan to report about our progress in our future publications.\\

\noindent
{\bf Competing Interests}\\
The authors declare that they have no competing interests.\\

\noindent
{\bf Acknowledgement:} One of us (TB) would like to gratefully acknowledge the financial 
support from BHU-fellowship  under which, the present investigation has been carried out.\\

\end{document}